\documentstyle[aps,epsf,psfig]{revtex}

\def\epsfig#1#2#3#4
         {
         \epsfysize=#2 \vbox{ \hglue#3 \epsfbox[#4]{#1} }
         }
\def\epsfigrot#1#2#3#4
         {
         \epsfxsize=#2
         \setbox\rotbox=\hbox to #2{\epsfbox[#4]{#1}}
         \vbox{\hglue#3 \rotl\rotbox}
         }
\newbox\rotbox
\begin{document}

\title{External voltage sources and Tunneling in quantum wires.}
\author{A. Koutouza, F. Siano and  H. Saleur$^\dagger$.}
\address{Department of Physics, University of Southern California,
Los-Angeles, CA 90089-0484.}
\address{$^\dagger$ CIT-USC Center for Theoretical Physics, USC, Los
Angeles, CA 90089-2535.}
\date{\today}
\maketitle

\begin{abstract}

We (re) consider in this paper  the problem of
tunneling through an impurity in a quantum
wire with  arbitrary Luttinger
interaction parameter. By combining the integrable approach developed in the case of
quantum Hall edge states with the introduction of radiative boundary conditions
to describe the adiabatic coupling to the reservoirs, we are able to obtain the exact
equilibrium and non equilibrium current. One of the most striking features observed is the
appearance of negative differential conductances out of equilibrium in the 
strongly interacting regime $g\leq .2$.   In spite of the various charging 
effects, a remarkable form of duality is still observed.

New results on the computation of transport properties in integrable
impurity problems are gathered in appendices. In particular, we prove that
the TBA results satisfy a remarkable relation, originally derived using the Keldysh formalism, between the order $T^2$
correction to the current out of equilibrium and the second derivative
of this  current at $T=0$  with respect to the voltage.
\end{abstract}

\section{Introduction.}

Tunneling experiments \cite{KF} are one of the most
efficient probes of the physics of  Luttinger liquids, which
 is expected to
describe the properties of one dimensional conductors. The case of spinless Luttinger liquids
has already been extensively studied, both theoretically and experimentally, in the
context of edge states in a fractional quantum Hall bar, where in particular, shot noise measurements
have led to the observation of fractional charge carriers \cite{noise}. The full cross-over  between
the weak and strong backscattering regimes has also been studied \cite{moon,milliken,FLS}: it exhibits in
particular  a duality between Laughlin quasi particles and electrons that is the result of the strong interactions
in the system, and, ultimately, of integrability. From a theoretical point of view,
it must  be stressed that crossovers in this type of problems
 can only be properly studied with
non perturbative methods anyway. In fact, for the physics out of equilibrium, which plays a crucial role in
the shot noise experiments for instance, numerical simulations  don't even seem to be available.

Other one dimensional conductors where Luttinger liquid physics could be observed
include carbon nanotubes \cite{nanotubes}, or quantum wires in semiconductor heterostructures \cite{quantumw}.
A key question for the latter examples is how to describe the application
of an external voltage. In the fractional quantum Hall case, this turned out to be easy \cite{FLS} because the
left and right moving excitations are physically {\sl separated} (the Luttinger liquid is really the
``sum'' of two independent chiral ones), and put at a different chemical potential
by the applied voltage. This will not be the case in a real quantum wire, where various
charging effects  have to be taken into
account.

The matter led to some active debating \cite{charging,EG}, and now seems quite
settled.  We
follow here the  approach of \cite{EG}, which easily allows
the inclusion of an impurity.  We thus  consider a gated quantum wire coupled adiabatically
to 2D or 3D reservoirs. As in Landauer-Buttiker's approach for non interacting electrons \cite{Landauer},\cite{Buttiker},
these reservoirs are assumed to be ``ideal'', and merely are there to inject bare densities of
left and right movers in the wire. The interactions in the wire lead to the appearance of
a non trivial electrostatic potential, and, in turn, to a renormalized charge density in the wire,
in the absence of impurity. When the impurity is present, there is in addition  a non trivial
spreading of the charges along the wire.

The key ingredient in the analysis of \cite{EG} is the equivalent of Poisson's equation,
which becomes a relation between the
electrostatic potential $\varphi $ and the charge density: $e\varphi =u_{0}\rho $.
Here, $u_{0}$\ is related to the Luttinger liquid constant by $
g=(1+u_{0}/\pi \ \hbar \ v_{F})^{-1/2}$. The electrostatic potential in turn shifts
the band bottom, and thus the total density. There follows a relation between
the bare injected densities and the true densities :

\begin{eqnarray}
\rho _{R}^{0} & = & \frac{g^{-2}+1}{2}\rho _{R}+\frac{g^{-2}-1}{2}\rho _{L}
\nonumber\\
\rho _{L}^{0} & = & \frac{g^{-2}-1}{2}\rho _{R}+\frac{g^{-2}+1}{2}\rho _{L}.
\label{initeq}
\end{eqnarray}

As for the bare densities themselves, they are related with  the external
voltage sources

\begin{eqnarray}
\rho _{R}^{0}(-L/2) & = & \frac{eU}{4\pi \hbar v_{F}} \nonumber\\
\rho _{L}^{0}(L/2) & = & -\frac{eU}{4\pi \hbar v_{F}}.\label{boucond}
\end{eqnarray}

The hamiltonian including the impurity term reads then, after bosonization 
\begin{equation}
H=\frac{\hbar v}{8\pi }\int dx\left[ \left( \partial _{x}\phi _{R}\right)
^{2}+\left( \partial _{x}\phi _{L}\right) ^{2}\right] +\lambda \cos \left[ 
\sqrt{g}\left( \phi _{R}-\phi _{L}\right) \right] (0),  \label{hamil}
\end{equation}
where $v=\frac{v_{F}}{g}$ is the sound velocity.

To proceed, one defines odd and even combinations of the bosonic field. Only
the even field interacts with the outside potential, and gives rise to a
current. Setting
\begin{equation}
\phi _{e,o}=\frac{1}{\sqrt{2}}\left[ \phi_{R}(x)\mp \phi_{L}(-x)\right],
\end{equation}
the hamiltonian of interest is
\begin{equation}
H_{e}=\frac{\hbar v}{8\pi }\int dx\left( \partial _{x}\phi _{e}\right)
^{2}+\lambda \cos \left( \sqrt{2g}\phi _{e}\right) (0),  \label{hamileven}
\end{equation}
where $\phi _{e}$ is a pure right moving field. In these new
variables, the boundary conditions (\ref{boucond}) read
\begin{equation}
\left( g^{-1}+1\right) \rho _{e}(-L/2)-\left( g^{-1}-1\right) \rho _{e}(L/2)=
\sqrt{g\over 2}\frac{eU}{\pi \hbar v_{F}}.  \label{bdry}
\end{equation}
In going from (\ref{initeq}) to (\ref{bdry}), the relation
$\rho_{R,L}={\sqrt{g}\over 4\pi} \partial_x\left[\phi_R-\phi_L\pm{1\over g}\left(\phi_R+\phi_L\right)\right]$, that follows from bosonization, has been used. We
also have defined   $\rho_{e,o}={1\over 2\pi}\partial_x\phi_{e,o}$.
Finally, a mistake in \cite{EG} was  corrected (see  \cite{EGi}).

Our goal is to compute the current $I$ flowing through the system as
a function of the applied voltage $U$. In \cite{EG}, this was accomplished in the case $g={1\over 2}$, using a mapping
on free fermions. In this paper, we shall solve the problem for general values of $g$ using integrability of
the boundary sine-Gordon model \cite{GZ},\cite{FSW}. This paper can be
considered as a sequel - and to some extent a correction -
to the work \cite{FLS,LSS}, where the charging effects were not yet fully
understood. It is also an extension of the short letter \cite{usall}.

\section{General formalism}

First, we  set $e=v=\hbar =1$ (so $v_F=g$). To treat the interaction term at $x=0$ in an integrable way, one needs to chose an
appropriate basis for the bulk, massless, right moving excitations, which obey  $e=p$.
For $g$ generic, the basic  excitations can be kinks or antikinks -- carrying a $
\rho _{e}$ charge equal to $\pm 1$ -- and breathers. In the following we
shall often restrict for technical simplicity to $g=\frac{1}{t}$, $t$ an integer.
There are then $t-2$ breathers. We shall parametrize the energy of the
excitations with rapidities, $e_{j}=m _{j}e^{\theta }$. Here $m_{j}$ is
a parameter with the dimension of a mass; for kink and antikink, $m _{\pm
}=\mu $, while for breathers, $m _{j}=2\mu \sin \frac{j\pi }{2(t-1)}$, $j=1,\ldots,t-2$.
The value of $\mu$ is of course of no importance since the theory is massless, and in the following
we shall simply set it equal to unity.
The massless excitations enjoy factorized scattering in the bulk. At
a
temperature $T$, and with a choice of chemical potentials,
they have densities
given by solutions of the thermodynamic Bethe ansatz equations, which we shall generically denote by $\sigma_j$ (not to
be confused with charge densities).

The key point is that these excitations  have also a factorized scattering through the impurity,
described by a transmission matrix $T_{\pm \pm }$. This matrix depends on
the ratio of the energy of incident particles to a characteristic energy
scale $T_{B}$. In the following, it is useful to parametrize $T_{B}=
e^{\theta _{B}}$. The modulus square of the transmission matrix have very simple
expressions; we recall that
\begin{equation}
|T_{++}|^2= {e^{2\left({1\over g}-1\right)(\theta-\theta_B)}\over 1+
  e^{2\left({1\over g}-1\right)(\theta-\theta_B)}}.
\end{equation}
Finally, we also recall how $T_B$ is related with the bare coupling $\lambda$
\cite{FLS}:
\begin{equation}
T_B=\left(2\sin\pi g\right)^{1\over 1-g}{\Gamma\left({g\over 2(1-g)}\right)\over
\sqrt{\pi}\Gamma\left({1\over 2(1-g)}\right)}
\left[\lambda\Gamma(1-g)/2\right]^{1/(1-g)} \label{barerel}.
\end{equation}
To proceed, we start by expressing the boundary conditions in terms of the
massless scattering description:
\begin{eqnarray}
\rho_e(L/2) & = & {1\over \sqrt{2g}} \int_{-\infty }^{\infty }\left( \sigma _{+}\left|
T_{++}\right| ^{2}+\left| T_{+-}\right| ^{2}\sigma _{-}-\left|
T_{--}\right| ^{2}\sigma _{-}-\left|T_{+-}
\right| ^{2}\sigma _{+}\right)
d\theta \nonumber\\
& = & {1\over \sqrt{2g}} \int_{-\infty }^{\infty }\left( \sigma _{+}-\sigma _{-}\right) \left(
\left| T_{++}\right| ^{2}-\left| T_{+-}\right| ^{2}\right) d\theta. \label{rhoplus}
\end{eqnarray}
Here, $\sigma_\pm$ are the densities of kinks and antikinks; one has $\sigma _{\pm }=nf_{\pm }$ where the pseudo
energies $\epsilon _{\pm }$ are equal and satisfy $n=\frac{1}{2\pi }\frac{d
}{d\theta }\epsilon _{\pm }$, $n=\sigma +\sigma ^{h}$ the total density of
states of kinks or antikinks (the factor ${1\over \sqrt{2g}}$ occurs because it is the electric charge
${1\over 2\pi}\int \partial_x\phi_e$ associated with the fundamental kinks of the problem). The $\epsilon $'s follow from the solution of
the TBA system of equations
\begin{equation}
\epsilon _{j}=T\sum_{k}N_{jk}\frac{s}{2\pi }\star \ln \left( 1+e^{\frac{
\epsilon _{k}-\mu _{k}}{T}}\right),  \label{tba}
\end{equation}
where $s(\theta )=\frac{t-1}{\cosh ((t-1)\theta )}$, $g={1\over t}$,
and  $N_{jk}$ is
the incidence matrix of the following TBA\ diagram

\bigskip
\noindent
\centerline{\hbox{\rlap{\raise28pt\hbox{$\hskip5.2cm\bigcirc\hskip.25cm +$}}
\rlap{\lower27pt\hbox{$\hskip5.1cm\bigcirc\hskip.3cm -$}}
\rlap{\raise15pt\hbox{$\hskip4.8cm\Big/$}}
\rlap{\lower14pt\hbox{$\hskip4.7cm\Big\backslash$}}
\rlap{\raise15pt\hbox{$1\hskip1cm 2\hskip1.3cm\hskip.4cm t-3$}}
$\bigcirc$------$\bigcirc$------$\bigcirc$------$\bigcirc$------$\bigcirc$\hskip.5cm $t-2$ }}

\bigskip

\noindent
The equations (\ref{tba}) have to be supplemented
by asymptotic conditions $\epsilon_j\approx m_j e^\theta$ as $\theta\rightarrow\infty$.
In (\ref{tba}), the chemical potential vanishes for all the breathers
which have no $U(1)$\ charge. For the kinks and antikinks, $\mu _{\pm }=\mp
\frac{W}{2}$, where $W$ has to be determined self-consistently (the logic
here is that the external potential and the temperature determine uniquely
the average densities everywhere in the quantum wire. As always in
macroscopic statistical mechanics, this can be described instead by a
distribution with fixed chemical potentials, which is exactly what the TBA\
allows one to handle. By $U(1)$ symmetry, it is known in advance that only
the kinks and antikinks have a non vanishing chemical potential $\mu _{\pm
}=\mp \frac{W}{2}$ ). The filling fractions read then
\begin{equation}
f_{\pm }=\frac{1}{1+e^{(\epsilon \mp W/2)/T}}.  \label{fill}
\end{equation}
The charge density on the left side of the impurity reads simply $\rho_e
(-L/2)={1\over \sqrt{2g}}\int (\sigma _{+}-\sigma _{-})d\theta $ (it can be simply
reexpressed in terms of $W$: $\rho_e(-L/2)=\sqrt{g\over 2}{W\over 2\pi}$), so the boundary conditions
equation (\ref{boucond}) reads
\begin{equation}
\int \left( \left| T_{++}\right| ^{2}+\frac{1}{g}\left| T_{+\
-}\right| ^{2}\right) (\sigma _{+}-\sigma _{-})d\theta =
\frac{U}{2\pi}.  \label{bdrygene}
\end{equation}
The other key equation in the solution of the problem
follows from the charge density drop across the barrier
\begin{equation}
\Delta \rho =\rho (x<0)-\rho
(x>0)=g\frac{V}{\pi }.\label{drop}
\end{equation}
Here, $\rho=\rho_R+\rho_L$, and  $V$
is the four terminal voltage (that is, the voltage difference measured by
weakly coupled reservoirs on either side of the impurity; it consists of an
electrostatic potential drop, plus an electrochemical
contribution). By following the previous transformations,
one finds that $\Delta \rho =\Delta (\sigma_+-\sigma_-)$, and
thus (\ref{drop}) reads,
\begin{equation}
\int \left| T_{+\ -}\right| ^{2}(\sigma _{+}-\sigma _{-})d\theta =
g\frac{V}{2\pi }.  \label{densdrop}
\end{equation}
Finally, the tunneling current $I=\frac{U-V}{2\pi }$ reads, from
(\ref{drop}) and (\ref{bdrygene})
\begin{equation}
I=\int \left| T_{++}\right| ^{2}(\sigma _{+}-\sigma _{-})d\theta.
\label{curr}
\end{equation}
If $T/T_{B}$ or $U/T_{B}$ are large (the high energy, or weak backscattering limit), the solution of (\ref{bdrygene}) is $
\int (\sigma _{+}-\sigma _{-})d\theta \approx \frac{U}{
2\pi }$, and thus from (\ref{curr}), $I\approx \int (\sigma
_{+}-\sigma _{-})d\theta =\frac{U}{2\pi }$. This, once physical units are
reinstated, reads $I=\frac{e^{2}}{h}U$, the expected formula for a spinless
quantum wire.

From the foregoing system of equations, it is now easy to deduce
the following identity giving the parameter $W$ in terms of the
physical voltage and current
\begin{equation}
U=2\pi \left(1-{1\over g}\right)I+W.  \label{classii}
\end{equation}
The following relation is also quite useful:
\begin{equation}
V=W-{2\pi\over g}I.
\end{equation}

\section{ Results}

The limit  $g\rightarrow 1$, which describes non interacting electrons, is very simple. In that
case indeed, the $T$ matrix elements become  rapidity independent, and the system
of equations can readily be solved to give $V=\left| T_{+\ -}\right| ^{2}U$,
$I=\left| T_{+\ +}\right| ^{2}\frac{U}{2\pi }$. Here, the transmission probability
is not trivial in general, since, as $g\rightarrow 1$, $\theta_B$ has to diverge to ensure a finite value
of the bare coupling  $\lambda$ \cite{FSW,CKLM}.

The system of equations determining $I$ can  also be solved easily
in the ``classical limit'' $g\rightarrow 0$, where
\cite{FS} (this is detailed some more in the appendix)
\begin{equation}
I\approx 2g ~{T\over 2\pi}~\frac{\sinh (W/2T)}{I_{iW/2\pi T}(2x)I_{-iW/2\pi T}(2x)}
,~~ x=\frac{T_{B}}{4T},  \label{classi}
\end{equation}
$I$ are the usual Bessel functions, and  $W$ follows from (\ref{classii}). 

Closed form results can also be obtained for $g={1\over 2}$ (see below); besides, except at $T=0$, one
has to resort to a numerical solution of the TBA equations.
To tackle the physics of this problem as  $g$ varies, we consider first the linear conductance at temperature $T$. In the limit
$U\rightarrow 0$, the foregoing system of equations can easily be solved
by linearization, giving rise to
\begin{equation}
G=\frac{1}{2\pi }\frac{\int \left| T_{+\ +}\right| ^{2}\frac{d}{d\theta }
\left( \frac{1}{1+e^{\epsilon /T}}\right) d\theta }{\int \left( \left|
T_{+\ +}\right| ^{2}+\frac{1}{g}\left| T_{+\ -}\right| ^{2}\right)
\frac{d}{d\theta }\left( \frac{1}{1+e^{\epsilon /T}}\right) d\theta }={1\over 2\pi}\left[
 G_0+\left(1-{1\over g}\right) G_0\right],
\label{linearcond}
\end{equation}
where $G_0$ is the linear conductance in the quantum Hall effect problem \cite{FLS} (the numerator of this equation).  One of
the roles of the denominator is to renormalize the conductance from $g$ to
unity in the UV\ region. In the case $g={1\over 2}$, equation
(\ref{linearcond})
can be evaluated in closed form to give
\begin{equation}
G={1\over 2\pi} {1-{T_B\over 2\pi T}\Psi'\left({1\over 2}+{T_B\over
      2\pi T}\right)
\over
 1+{T_B\over 2\pi T}\Psi'\left({1\over 2}+{T_B\over
      2\pi T}\right)},
\end{equation}
where $\Psi$ is the digamma function. For values $g={1\over t}$,
$t$ an integer, $G$ is easily determined numerically by solving the
system
of TBA equations (\ref{tba}), and plotting the soliton pseudo energy
back into (\ref{linearcond}). Curves for various values of $g$ are shown
 in Fig.~1; features entirely similar to those
in \cite{FLS} are observed, although all the curves now converge to the same value in the high temperature limit, in contrast with the
quantum Hall edges case. The effect of the impurity is considerably amplified as $g$ gets smaller, with $G$ getting a discontinuity in the weak back scattering limit as $g\to 0$.
Indeed, letting  $U\to 0$ in (\ref{classii}), one finds
\begin{equation}
G={I\over U}\approx\frac{1}{2\pi }\frac{1}{\left(1-{1\over g}\right)+{1\over g}I_{0}^{2}(2x)}.
\end{equation}
As $g\to 0$, $G$ thus becomes a step function, jumping from ${1\over 2\pi}$ to $0$ as soon as $T_B$ ($T_B=2\lambda$ for $g=0$) is turned on, for any temperature.

\begin{figure}[h]
\centerline{\psfig{file=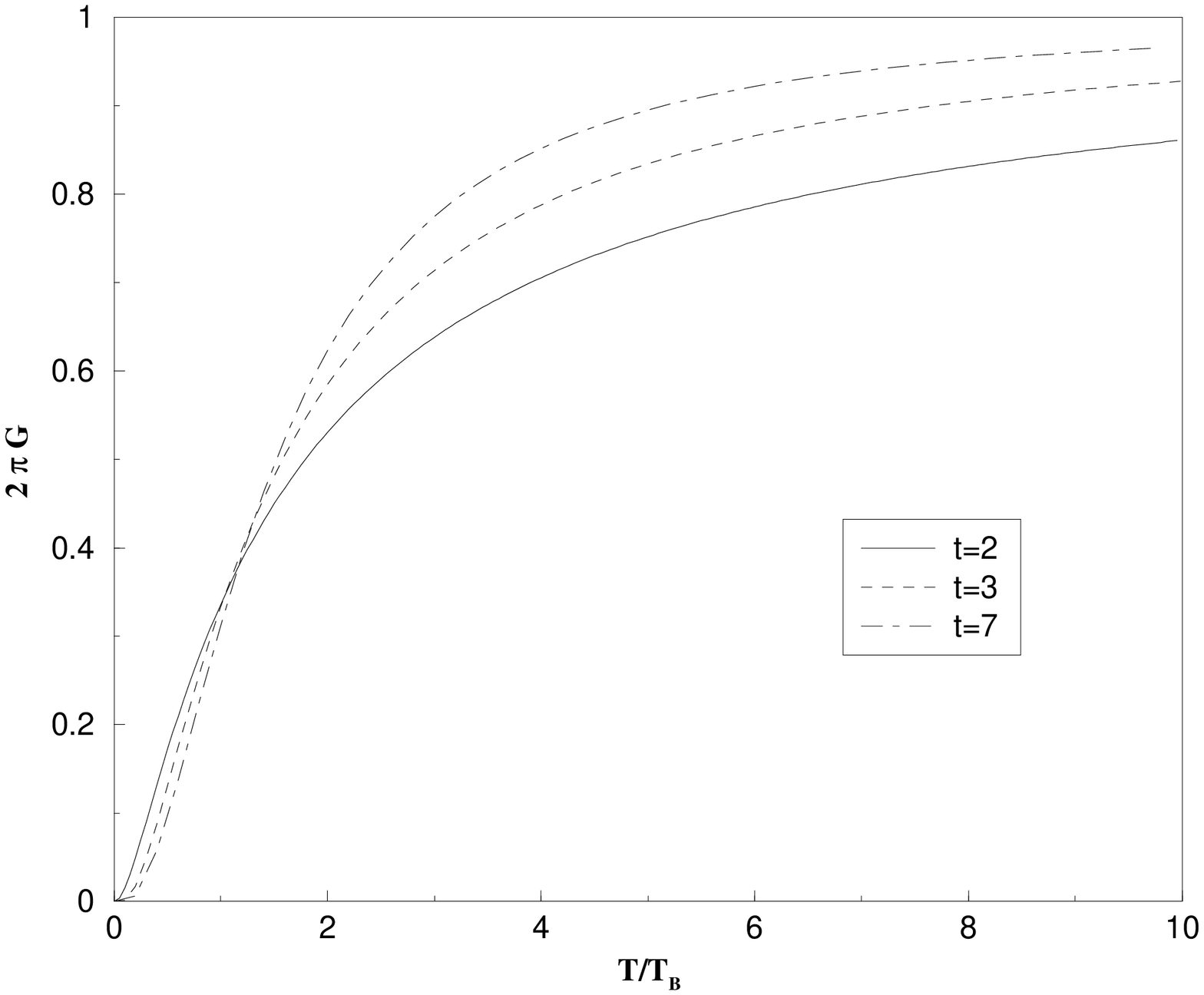,height=2.5in,angle=270}}
    \label{fig:plotofcond}
    \caption{We represent here the conductance as a function of the universal ratio
 of  temperatures $T/T_B$ for several values of $g=1/t$, $t$ an integer. In this domain - which is the
easiest to study numerically - $G$ has only a weak dependence on $g$. These curves interpolate between  two limiting behaviours: for $g=1$, $2\pi G$ should become a constant equal to $1/2$, while for $g=0$ $2\pi G$ should vanish for any finite value of $T/T_B$.  }
\end{figure}

Another simple limit to study is the case  $T=0$, where
results are  far more intriguing. Consider first the classical limit: as  $W$ is sweeped, one finds that $I$ vanishes while $U$ increases
up to  $\pi T_B$, then goes back to zero, beyond which $I$ increases like $I={U\over 2\pi}$. In other words, the system behaves either like a perfect insulator, or like a perfect conductor! This
very singular behavior is  the $g\to 0$ limit of a multivalued $I-U$ characteristics
with regions of negative differential conductance \cite{usall}, that we now study in more details.

\bigskip

Indeed, the TBA equations can   be solved in closed form in the limit $T\rightarrow 0$
. In that limit, $\sigma _{-}=0$, the integrals run only from $-\infty $ to
a cut off (Fermi) rapidity $A$, and $\sigma _{+}=n$ follows from the
solution of the integral equation
\begin{equation}
n(\theta )-\int_{-\infty }^{A}\Phi (\theta -\theta ^{\prime })n(\theta
^{\prime })d\theta ^{\prime }=\frac{e^{\theta }}{2\pi },  \label{intn}
\end{equation}
while $A$ is determined by the condition that $\epsilon_+^h (A)=0$, where 
\begin{equation}
\epsilon_+^h (\theta )-\int_{-\infty }^{A}\Phi (\theta -\theta ^{\prime
})\epsilon_+^h(\theta ^{\prime })d\theta ^{\prime }=\frac{W}{2}-e^{\theta }.
\label{inteps}
\end{equation}
In that equation, $\Phi$ is the derivative of the log of the kink kink $S$ matrix
$$
\Phi(\theta)=\int_{-\infty}^\infty e^{-i\omega\theta} {\sinh \pi\left({2g-1\over 2(1-g)}\right)\omega\over
2\cosh {\pi\omega\over 2}\sinh {\pi g\omega\over 2(1-g)}}{d\omega\over 2\pi}.
$$
Since $W$ determines $A$ uniquely (one finds $e^{A}=\frac{W}{2}\frac{
G_{+}(0)}{G_{+}(i)}$, where the propagators $G$ are defined below), in what follows we will consider instead $A$ as the
unknown when $T=0$. After a few rearrangements, the relevant equations read
now (we still set $g={1\over t}$, although $t$ does not have to be an integer here))
\begin{equation}
\int_{-\infty }^{A}n(\theta )~\frac{t+e^{2(t-1)(\theta -\theta _{B})}}{
1+e^{2(t-1)(\theta -\theta _{B})}}~d\theta ={U\over 2\pi},
\label{tzeroi}
\end{equation}
and
\begin{equation}
I=\int_{-\infty}^A n(\theta )~\frac{e^{2(t-1)(\theta -\theta _{B})}}{
1+e^{2(t-1)(\theta -\theta _{B})}}~d\theta.   \label{tzeroii}
\end{equation}
The density $n(\theta )$ can be computed as a power series in the weak and
strong backscattering limits, giving rise to expansions for the current and
the boundary conditions. In the strong backscattering case one finds :
\begin{equation}
I={G_{+}(i)\over G_+(0)}{e^A\over\pi}\sum_{n=1}^{\infty }(-1)^{n+1}\frac{\sqrt{\pi }
~\Gamma (nt)}{2\Gamma (n)\Gamma \left(\frac{3}{2}+n(t-1)\right)}\left( e^{A+\Delta
-\theta _{B}}\right) ^{2n(t-1)},  \label{sbsi}
\end{equation}
while the boundary condition reads
\begin{equation}
2{G_{+}(i)\over G_+(0)}e^A-(t-1){G_{+}(i)\over G_+(0)}e^A\sum_{n=1}^{\infty }(-1)^{n+1}\frac{
\sqrt{\pi }~\Gamma (nt)}{\Gamma (n)\Gamma \left(\frac{3}{2}+n(t-1)\right)}\left(
e^{A+\Delta -\theta _{B}}\right) ^{2n(t-1)}=U.  \label{bdri}
\end{equation}
In the weak backscattering limit instead, one finds
\begin{equation}
I={G_{+}(i)\over  G_+(0)}{e^A\over \pi t^{2}}\left[ t-\sum_{n=1}^{\infty }(-1)^{n+1}
\frac{\sqrt{\pi }~\Gamma (n/t)}{2\Gamma (n)\Gamma \left(\frac{3}{2}+n(\frac{1}{t
}-1)\right)}
\left( e^{A+\Delta -\theta _{B}}\right) ^{2n(\frac{1}{t}-1)}\right],
\label{wbsi}
\end{equation}
where
\begin{equation}
\frac{2}{t}{G_{+}(i)\over G_+(0)}e^A-{1\over t}\left({1\over t}-1\right){G_{+}(i)\over G_+(0)}e^A\sum_{n=1}^{
\infty }(-1)^{n+1}\frac{\sqrt{\pi }~\Gamma (n/t)}{\Gamma (n)\Gamma \left(\frac{3
}{2}+n(\frac{1}{t}-1)\right)}\left( e^{A+\Delta -\theta _{B}}\right) ^{2n\left(\frac{1
}{t}-1\right)}=U.  \label{bdrii}
\end{equation}
Here we have introduced the notations
\begin{equation}
G_{+}(\omega )=\sqrt{2\pi t}\frac{\Gamma \left( -i\omega \frac{t}{2(t-1)}
\right) }{\Gamma \left( \frac{1}{2}-i\frac{\omega }{2}\right) \Gamma
\left( -i\omega \frac{1}{2(t-1)}\right) }e^{-i\omega \Delta },  \label{prop}
\end{equation}
where
\begin{equation}
\Delta =\frac{1}{2}\ln (t-1)-\frac{t}{2(t-1)}\ln t.  \label{deltt}
\end{equation}
In terms of the auxiliary variable $W$, the strong and weak backscattering
expansions have matching radius of convergence: either one of them is
always converging, and both are  at the matching value $\frac{W}{
T_{B}^{\prime }}e^{-\Delta}=1$, where the parameter $T_{B}^{\prime }$ is defined as
\begin{equation}
T_{B}^{\prime }=2T_{B}e^{-\Delta }\frac{G_{+}(i)}{G_{+}(0)}.  \label{param}
\end{equation}

The series can be summed up in the case $g={1\over 2}$ to give
\begin{equation}
\tan {U-2\pi I\over 2}={U+2\pi I\over 2}.
\end{equation}

There is a rich physical behavior hidden in these equations. To investigate it,
consider first the behavior of physical quantities as a function of $W$. Curves representing $U$ and
${2\pi\over g} I$ as a function of $W$ for various $t$ are given in Fig.~2
 and Fig.~3.

\begin{figure}[h]
\centerline{\psfig{file=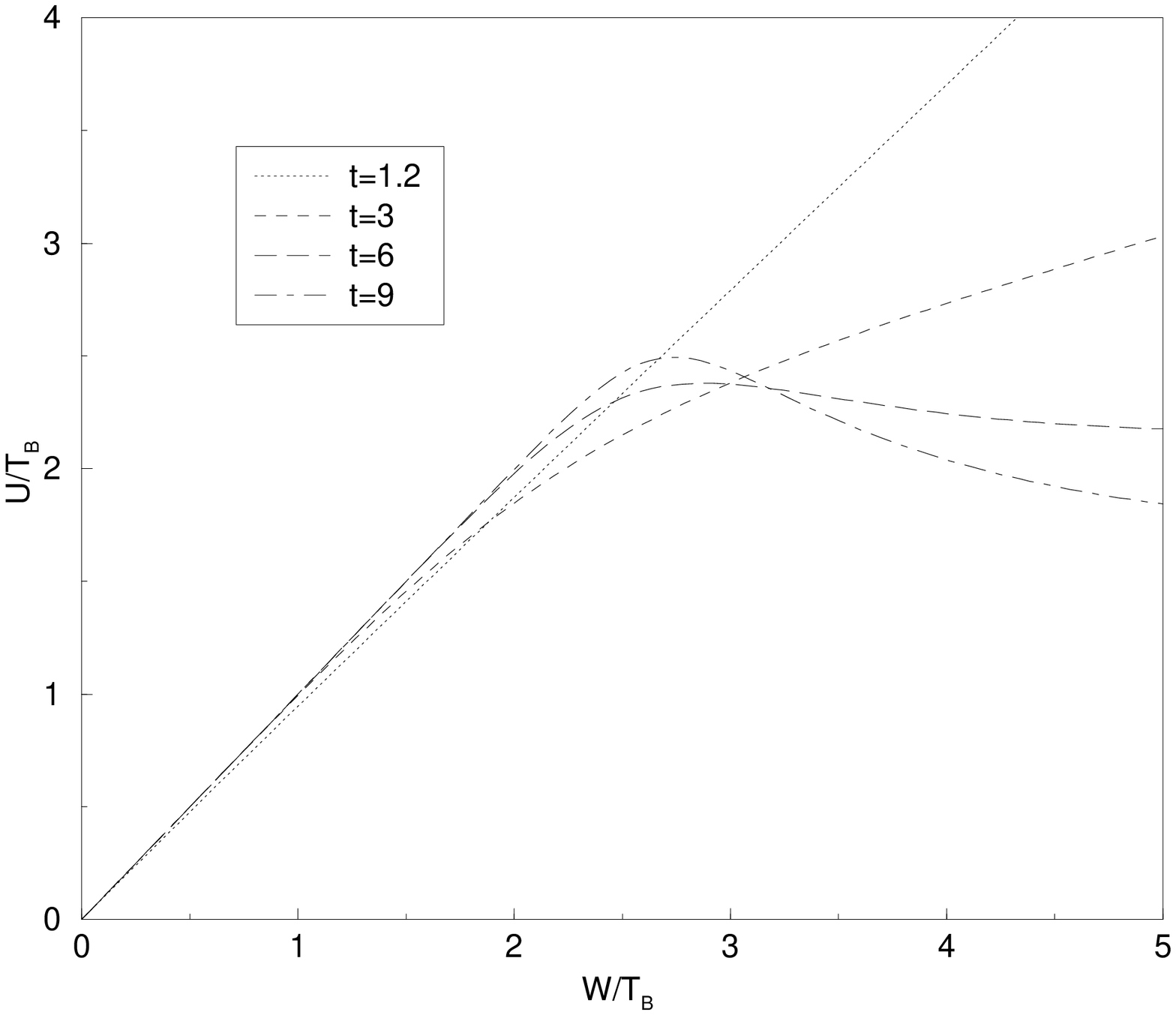,height=2.5in,angle=270}}
   \label{UW}
   \caption{The applied voltage difference $U/T_B$ as a function of the chemical potential difference between solitons and antisolitons, $W/T_B$. Observe the remarkable non monotonic behaviour that settles in for small enough values of $g$. This results in the existence of two possible values of $W$ for a given $U$, and thus in the existence of the loop in the $I-U$ characteristic.}
\end{figure}

\begin{figure}[h]
\centerline{\psfig{file=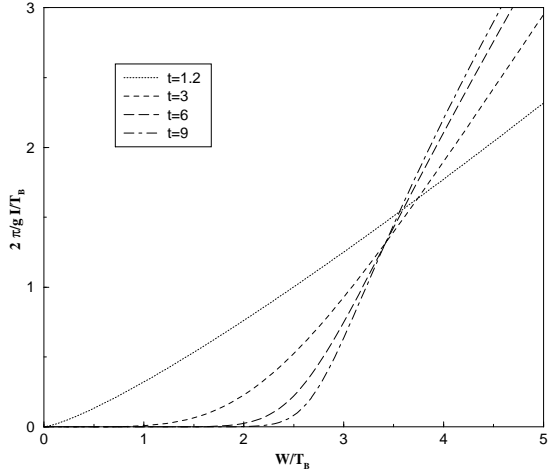,height=2.5in,angle=270}}
   \label{T0IW}
   \caption{In contrast, the current $I$ as a function of $W$ exhibits, once properly rescaled, a very weak dependence on $g$. All curves behave asymptotically as $W/T_B$ in the weak backscattering limit. }
\end{figure}

As $g\to 0$, the current in the strong backscattering expansion
is exactly $0$. In the weak backscattering expansion meanwhile, it reads
$$
{2\pi I\over g}\approx\left(W^2-\pi^2 T_B^2\right)^{1/2}
$$
hence exhibits a square root singularity at finite value of $W$ (we note
that the latter expression can also be obtained directly
from the result (\ref{classi}) by using the uniform asymptotic expansion
of Bessel functions for large orders \cite{AZ}:
$$
I_\nu(\mu z)\approx {1\over \sqrt{2\pi\nu}} {e^{\nu\eta}\over (1+z^2)^{1/4}},
$$
where $\eta=\sqrt{1+z^2}+\ln {z\over 1+\sqrt{1+z^2}}$). When $t$ is varied,
the current evolves from
 this singular behavior to the  simple characteristics
$I={W\over 4\pi}$ as $g\to 1$  (this is easily seen from the integral representations of $I$ and $U$:
and an artifact of the variable $T_B$ used throughout, that would have to
be rescaled appropriately in that limit to give a non trivial $I-U$ relation \cite{FSW}). At fixed $g\neq 1$, $I\approx {W\over 2\pi}$ at large $W$.

As $g\to 0$, $U$ in the strong backscattering expansion is simply equal to $W$, while in
the weak backscattering expansion it reads
$$
U\approx W-\left(W^2-\pi^2 T_B^2\right)^{1/2}.
$$
As $g\to 1$ meanwhile, $U\approx W$. When $g$ varies, $U$ interpolates between these two
limiting behaviors, and stops having a (local) maximum around $t\approx 4.83$.

The fact that $U$ can {\sl decrease} as $W$ increases is a direct consequence of the physics in this system. The
density on the left, $\rho_e(-L/2)\propto W$. An increase in $W$ increases the left density,
but it also increases the right density, since particles being more energetic, more of them go across the impurity. $U$ is a non trivial function of the densities on either side of the impurity,
as given by (\ref{bdry}). For $g$ large, $U$ behaves essentially as the sum
of the densities in $\pm L/2$, thus increases when $W$ increases. However, when $g\to 0$, $U$
gets dominated by the difference of the densities, and if enough particles go across, it can
well decrease when $W$ increases. This effect is directly related to the fact that
the differential conductance $2\pi {dI\over dW}$ does, for $g<{1\over 2}$,
 actually get {\sl larger} than $g$ for finite
values of $W$  an effect first observed in \cite{FLS} (see Fig.~4).

\begin{figure}[h]
  \centerline{\psfig{file=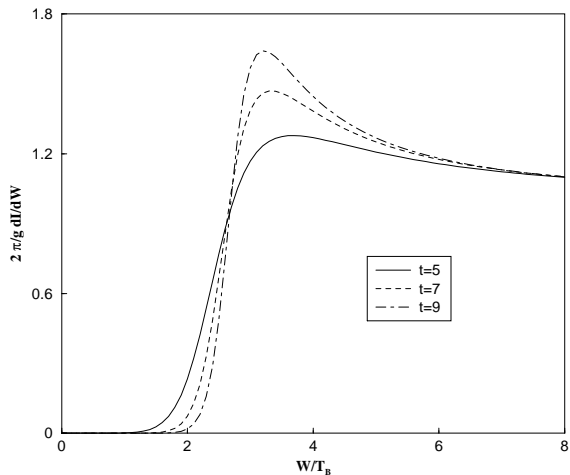,height=2.5in,angle=270}}
  \label{andreI}
  \caption{The rescaled derivative of the current with respect to $W$ at $T=0$. Notice the existence of a maximum above the weak backscattering limit (equal to $1$) for $t\geq 2$. This peak of differential conductance becomes more and more marked as $g\rightarrow 0$. }
\end{figure}

Consider now $I$ as a function of $U$: clearly, the existence of a
maximum in the curve $U(W)$
will lead to an S-shaped $I(U)$. More precisely, consider first the case
$g\approx 0$. Suppose we increase $W$ starting from $0$. According to
Fig.~2,$U$ first increases up to $\pi T_B$, then decreases
back to zero.  $W$ being still finite, $I$ vanishes identically, since it has
an overall factor of $g$. Going now to the regime where $W$ becomes infinite,
$U\approx W$, and $I\approx {W\over 2\pi}\approx {U\over 2\pi}$: the system has switched from being a perfect insulator to
being a perfect conductor! This is easy to understand in more
physical terms: as $g\to 0$, the kinetic term dominates the Lagrangian,
and one might expect that the impurity is essentially invisible. However, as $g\to 0$,
there is the possibility that a charge density wave might form, getting pinned
down by an infinitesimal potential, and leading to a perfect insulator \cite{usall}.

This effect is stable against quantum fluctuations, and   for $g$ approximately smaller than $g=.2$, a ``loop'' keeps being observed in the $I-U$ characteristics. That the
 current is not a single valued function of $U$ in the region of small voltages, leads  to the
prediction of  hysteresis and bistability in the strongly interacting, out of
equilibrium regime. Although the present calculation is valid only in the scaling regime, this qualitative aspect should survive beyond it.

The loop is also stable against thermal fluctuations: as is illustrated in 
Fig.~5 for the case $t=6$,
it only disappears at a finite temperature $T_c$ which depends on $g$.

\begin{figure}[h]
\centerline{\psfig{file=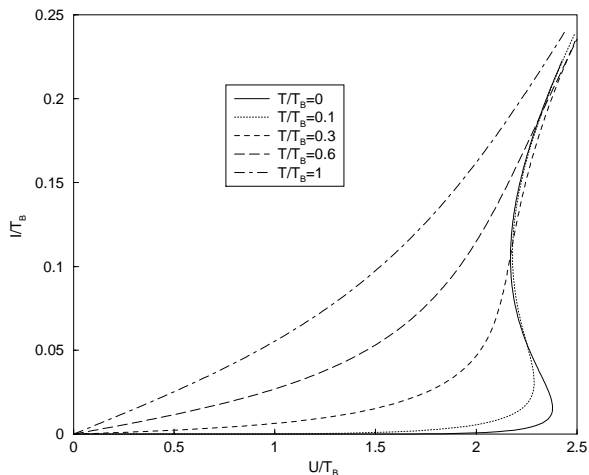,height=2.5in,angle=270}}
   \label{loopdis}
   \caption{We illustrate on this figure the disappearance of the $S$ shape as the temperature is increased. Clearly, the bistability is stable against
thermal fluctuations in a finite range which depends on $g$. Here, $g=1/6$. }
\end{figure}

A semi classical approximation \cite{usall} gives
$T_c=T_B \sqrt{(1-g)\over 16 g}$: this formula is not quite correct for values of $g\leq .2$, but
becomes increasingly good as $g\to 0$. It is quite difficult numerically to determine
$T_c$ with a good accuracy: a reasonable estimate of this curve is given in Fig.~6.

\begin{figure}[h]
\centerline{\psfig{file=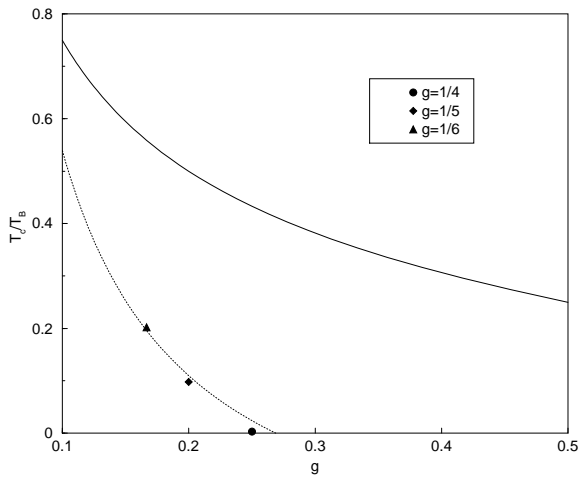,height=2.5in,angle=270}}
   \label{tempcrit}
   \caption{The ``critical'' temperature $T_c(g)$ at which bistability
disappears. Notice the poor quality of the leading semi-classical approximation (full curve).}
\end{figure}

\section{Duality}

For the problem of tunneling between quantum Hall edges, a striking duality
between the weak and strong backscattering limits was uncovered in \cite{FLS} at $T=0$, and further generalized to any $T$ \cite{BLZ}. The meaning of this duality was that, while the hamiltonian
describing the vicinity of the weak backscattering limit is given by (\ref{hamileven}),
the one describing the vicinity of the strong backscattering limit can be reduced, as far as the DC current is concerned, to an expression identical with (\ref{hamileven}), up to the replacement
of the coupling $\lambda$ by a dual coupling $\lambda_d$, together
with the exchange $g\rightarrow
 {1\over g}$. As a result, a duality relation for the current followed
\begin{equation}
I\left(\lambda,U,g\right)={gU\over 2\pi}-g I\left(\lambda_d,gU,{1\over g}\right)\label{olddual}.
\end{equation}
Here,  the dual coupling $\lambda_d$ reads
\begin{equation}
\lambda_d={1\over \pi g}\Gamma\left({1\over g}\right)
\left[{g\Gamma(g)\over \pi}\right]^{1\over g}
\lambda^{-{1\over g}}. \label{durel}
\end{equation}
The relation (\ref{durel})
follows from keeping the parameter
\begin{equation}
T_B''\equiv {T_B' \over \sqrt{t}},
\end{equation}
constant\footnote{In  \cite{FLS}, the duality
relation was initially written at constant $T_B'$. While the
identities
in \cite{FLS} are algebraically correct, it is really $T_B''$ that has
to be kept constant, since the applied voltage is not left invariant
in the duality transformation.} while letting $g\to {1\over g}$, and
using the relation (\ref{barerel}) between $T_B$ and the bare coupling
in the tunneling hamiltonian.

For pedagogical purposes, it is probably wise to explain a little more
explicitly what the duality means. Consider thus a hypothetical current
defined non perturbatively
by the expression
\begin{equation}
I={1\over x^2+g^2}.
\end{equation}
It obeys the following duality relation
\begin{equation}
I\left({1\over x},{1\over g}\right)=g^2-g^4 I(x,g).\label{exampdual}
\end{equation}
Suppose now we did not know the non perturbative expression, but had only access to
the small x expansion
\begin{equation}
I={1\over g^2}\sum_{n=0}^\infty (-1)^n \left({x^2\over g^2}\right)^n,
\end{equation}
and the large x one
\begin{equation}
I={1\over x^2}\sum_{n=0}^\infty (-1)^n \left({g^2\over x^2}\right)^n.
\end{equation}
The duality (\ref{exampdual}) could then be deduced from the expansions by say starting from
the small $x$ one, setting $x={1\over x'},g={1\over g'}$, and comparing the new
expression with the large $x$ expansion. What was done in \cite{FLS} was to find a similar
duality only based on the weak and strong backscattering expansions (a non perturbative
expression for the current was found much later \cite{duality}).

It is interesting to examine what
does remain of this duality in the present case. The IR hamiltonian
will behave similarly to the case of tunneling between quantum Hall
edges,
since it is entirely determined by the large $\lambda$ behavior, and
has no relation with the way the voltage is taken into account. This
means
that the parameter $T_B''$ still has to be kept constant in whatever
duality
symmetry one is looking for.

There is a quick way to proceed assuming from \cite{FLS}  the relation
(\ref{olddual}), which  becomes here
\begin{equation}
I(\lambda,W,g)={gW\over 2\pi} - gI\left(\lambda_d,gW, {1\over g}\right).\label{oldduali}
\end{equation}
Using this, together with the relation (\ref{classii}), one finds the
additional relation
\begin{equation}
U(\lambda, W,g)=U\left(\lambda_d, gW,{1\over g}\right).\label{olddualii}
\end{equation}
From this it follows that
\begin{equation}
I\left(\lambda,U,g\right)={U\over 2\pi}- I
\left(\lambda_d,U,{1\over g}\right).\label{newduality}
\end{equation}

For completeness, we can also give a direct proof of this relation.
It is convenient first
to put the equations in a more compact form, namely
\begin{eqnarray}
\Lambda_s\left[ 1-(t-1)\sum\limits_{n=1}^{\infty }\alpha
_{n}\Lambda_s^{2n(t-1)}\right] =u_s \nonumber\\
i_s(t,u_s)=\Lambda_s\sum\limits_{n=1}^{\infty }\alpha
_{n}\Lambda_s^{2n(t-1)},   \label{stbali}
\end{eqnarray}
for the strong backscattering limit, and
\begin{eqnarray}
{\Lambda_w\over t}\left[ 1-\left( \frac{1}{t}-1\right)
\sum\limits_{n=1}^{\infty }\beta _{n}\Lambda_w^{2n(
\frac{1}{t}-1)}\right] =u_w \nonumber\\
i_w(t,u_w)=\frac{1}{t^{2}}\Lambda_w\left[ t-\sum\limits_{n=1}^{\infty }\beta
_{n}\Lambda_w^{2n(\frac{1}{t}-1)}\right], \label{webali}
\end{eqnarray}
for the weak backscattering limit. In (\ref{stbali}),(\ref{webali})
\begin{eqnarray}
\alpha _{n} & = & \frac{\sqrt{\pi }}{2}(-1)^{n+1}2^{n(t-1)}\frac{\Gamma
(nt)}{\Gamma (n)\Gamma (\frac{3}{2}+n(t-1))} \nonumber\\
\beta _{n} & = & \frac{\sqrt{\pi }}{2}(-1)^{n+1}2^{n\left( \frac{1}{t}
-1\right) }\frac{\Gamma (n/t)}{\Gamma (n)\Gamma (\frac{3}{2}+n(\frac{1}{t}
-1))}=\alpha_n\left({1\over t}\right) \nonumber\\
i_s=i_w  =  \sqrt{2\over t}{\pi I\over T_B''} & , & u_s=u_w= \sqrt{2\over t}{U\over 2 T_B''}.\\
\end{eqnarray}
To match with our previous notations, $\Lambda=G_+(i,t) {e^A\over T_B''}$; however, $\Lambda$ in the foregoing
equations is determined by the external voltage, and no reference to $e^A$ or $T_B''$ are necessary in its definition.

It follows from (\ref{stbali}) and (\ref{webali})  that
\begin{equation}
i_s\left(t,u\right)={1\over t}\Lambda\left({1\over t},tu_w\right)
-{1\over t^2} i_w\left({1\over t},tu\right)\label{naivedual},
\end{equation}
where the parameter $\Lambda$ is the same in both $i_s$ and $i_w$. Of course, the current is an analytical function of the applied
voltage, independent of whether one considers the weak or strong
backscattering expansions, so the labels $s,w$ can actually be suppressed
from the equations. It follows that, going back to physical variables,
\begin{equation}
I\left(\lambda,U,g\right)= {1\over 2\pi} W\left({1\over g},
  U\right) - g I\left(\lambda_d,U,{1\over g}\right).
\end{equation}
Now, $W$ in turn can be expressed in terms of $U,I$, using the
relation (\ref{classii}), reproducing (\ref{newduality}).

The relation between the current and the applied voltage is implicit
in the
foregoing equations. It can, however, be made explicit by elimination
of the parameter $\Lambda$,
and we quote here the lowest orders for completeness.  In the weak backscattering limit one has
\begin{eqnarray}
i & = & -{1\over t}\beta _{1}(tu)^{2(\frac{1
}{t}-1)+1}- \nonumber\\
&  & \left[\frac{1}{t^{3}}(t-1)(t-2)\beta _{1}^{2}-{1\over t}\beta _{2}
\right](tu)^{4(\frac{1}{t}-1)+1}+... \\
\end{eqnarray}
and in the strong backscattering limit
\begin{eqnarray}
i & = & \alpha _{1}u
^{2(t-1)+1}+ \nonumber\\
&  & \left[(t-1)(2t-1)\alpha _{1}^{2}+\alpha
_{2}\right]u^{4(t-1)+1}+... \\
\end{eqnarray}
Meanwhile, the parameter $\Lambda$ can also be expanded, say in the
weak backscattering limit:
\begin{eqnarray}
\Lambda & = & \left({1\over t}-1\right)\beta _{1}(tu) ^{2(t-1)+1}+ \nonumber\\
&  & \left[ (t-1)(2t-1)\alpha_{1}^{2}+\alpha _{2}\right] u^{4(t-1)+1}+...
\end{eqnarray}
One can directly check the duality relation (\ref{naivedual}) on these
formulas. Notice  that despite the more complex physics, which 
now involves screening,
the exponents of the weak and strong backscattering expansions are 
the same
than in the fractional quantum Hall case. 

Finally, the duality was extended to finite temperatures in
\cite{BLZ}, \cite{LS}, meaning that formula (\ref{olddual}) holds
at finite temperature. Since (\ref{classii}) is still true too,
 the formula (\ref{newduality}) extends to finite temperature
as well.

\section{Conclusions}

This paper hopefully solves the tunneling problem with a proper treatment of the coupling to the reservoirs, hence completing and
correcting \cite{FLS,LSS}. We have only treated here the spinless case,
but the method extends straightforwardly to the spinfull case,
at least when the spin isotropy is not broken, and the problem maps onto a super symmetric boundary
sine-Gordon model \cite{LSS}

The duality we observed
does raise interesting physical questions, in particular concerning
the nature of the ``charges'' that tunnel in the weak backscattering limit. We hope to get back to this issue  with
computations of the DC shot noise.

\bigskip

\noindent {\bf Acknowledgments}: We thank R. Egger and H. Grabert
for an earlier collaboration on part of this material, for communicating
the results of \cite{EGi} before publication, and for many illuminating
discussions.  This work was supported by the DOE and the NSF (under the NYI program).

\appendix\section{Semi-classical computations}

In studying the classical limit, one usually concentrates on the behavior of $\epsilon_j$ for $j$ finite
while $g\rightarrow 0$, that is $t\rightarrow\infty$ \cite{Fowler}. This is not sufficient in the study of transport properties,
where the knowledge of $\epsilon_\pm$, that is pseudoenergies for nodes at the other end of the diagram,
are required.  The necessary analysis is a bit more complicated then. First,
it is convenient to introduce the new quantity $Y_j(\theta)\equiv e^{\epsilon_j(\theta)/T}$, and
to recast the TBA system, using the identity $s\left(\theta+{i\pi\over 2(t-1)}\right)
+s\left(\theta-{i\pi\over 2(t-1)}\right)= 2\pi\delta(\theta)$, into
\begin{equation}
Y_j\left(\theta+{i\pi\over 2(t-1)}\right)Y_j\left(\theta-{i\pi\over 2(t-1)}\right)=\left[1+Y_{j+1}(\theta)\right]
\left[1+Y_{j-1}(\theta)\right]\label{functba}
\end{equation}
In the limit where $g\rightarrow 0$, we introduce new variables $s\equiv {j\over t}$, $\alpha={2\theta\over\pi}$,
and $e^{-\chi}\equiv {e^\epsilon\over t^2}$,
and expand the left and right hand sides of equation (\ref{functba}) to obtain the Liouville equation \cite{wiegmann}
\begin{equation}
\left(\partial_s^2+\partial^2_\alpha\right) \chi=2 e^\chi \label{conttba}
\end{equation}
The general solution of this equation that is relevant here is \footnote{This is of the general
form of solution $e^{-\chi}={(1-A(z)B(\bar{z}))^2\over
4\partial A \bar{\partial B}}$ for the
equation $\partial\bar{\partial} \chi={1\over 2} e^\chi$,
 where $A={J_{-\rho}\over J_\rho}(e^z)$, $B={J_\rho\over
   J_{-\rho}}(e^{\bar{z}})$.}
\begin{equation}
e^{-\chi}={1\over \left( 2i\sin\pi\rho\right)^2}\left\{J_\rho\left[e^{{\pi\over 2}(\alpha+is)-\ln (2T)}
\right]J_{-\rho}\left[e^{{\pi\over 2}(\alpha-is)-ln(2T)}\right]-
(\rho\to -\rho)\right\}^2\label{bess}
\end{equation}
where $J_\rho$ are the usual Bessel functions, $\rho={iW\over 2\pi T}$. The freedom in the arguments of the Bessel functions
$\alpha+is\to \lambda(\alpha-\alpha_o+i(s-s_0))$
has been resolved by matching with  the asymptotic boundary conditions $\epsilon_j\approx 2 \sin\left({j\pi\over 2(t-1)}\right)
e^\theta$ as $\theta\rightarrow\infty$. As for the index of the Bessel functions, it is obtained by matching against the result at low energies:
$$
e^{\epsilon_j(-\infty)/T}=\left[{\sinh (j+1)W/2tT\over\sinh W/2tT}\right]^2-1
$$
We can now compute $\epsilon_{t-2}$ by setting $s=1$ in the solution (\ref{bess}): one finds
$$
e^{-\chi(\alpha,1)}=\left[J_\rho J_{-\rho}\left(i e^{{\pi\over 2}\alpha-\ln (2T)}\right)\right]^2
$$
It follows that
\begin{equation}
e^{\epsilon_\pm(\theta)/T}=t I_\rho\left({e^\theta\over 2T}\right)I_{-\rho}\left({e^\theta\over 2T}\right)\label{final}
\end{equation}
The current on the other hand reads
\begin{eqnarray}
I&=\int_{-\infty}^\infty |T_{++}|^2 \left(\sigma_+-\sigma_-\right)d\theta\nonumber
&={T\over 2\pi} \int_{-\infty}^\infty d\theta {1\over 1+e^{-2(t-1)(\theta-\theta_B)}}{d\over d\theta}
\ln {1+e^{-W/2T}e^{-\epsilon/T}\over 1+e^{W/2T}e^{-\epsilon/T}}
\end{eqnarray}
In the limit $t\rightarrow\infty$, this becomes then
\begin{equation}
I={T\over 2\pi} 2\sinh (W/2T) e^{-\epsilon_\pm(\theta_B)/T}
\end{equation}
Replacing $\epsilon_\pm$ by his classical expression reproduces then
the result (\ref{classi}).

\section{Low temperature expansion.}

The remarkable relation \cite{Weiss}
\begin{equation}
I(W,T)=I(W,T=0)+{\pi^2 T^2 t\over 3} {d^2 I\over dW^2}(W,T=0)\label{weissrel}
\end{equation}
was initially discovered, following a Keldysh expansion of the left and right hand sides, in the context of dissipative quantum mechanics in \cite{Weissetal}.
In (\ref{weissrel}), $W$ is the chemical potential defined in the text - it
would coincide with the Hall voltage $V$ in the context of the fractional
quantum Hall effect \cite{FLS}.

\begin{figure}[h]
\centerline{\psfig{file=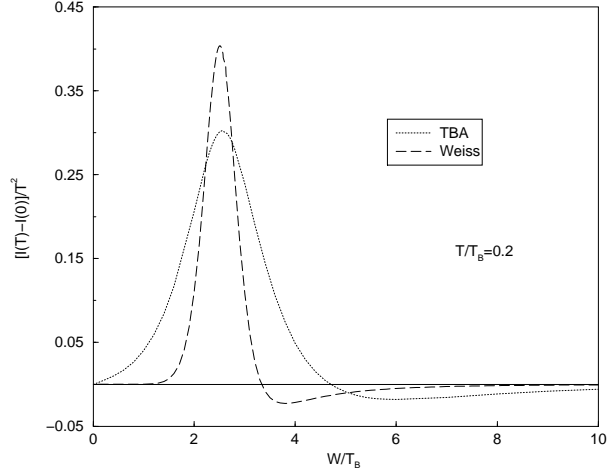,height=2.5in,angle=270}}
   \label{andreII}
   \caption{The dashed line is  the order $T^2$ correction to the non
equilibrium current as estimated by the equation (\ref{weissrel}). The dotted line is the same correction calculated from the TBA at $T=0.2$ (it is difficult, for technical reasons, to go below this value with enough accuracy). The two curves are in good qualitative agreement: notice that both of them are below the axis in the weak backscattering limit. On this figure, $t=7$. }
\end{figure}

We shall now prove  that the current obtained from the TBA does satisfy this relation indeed: as (\ref{weissrel}) involves out of equilibrium quantities {\sl and} the temperature, it provides a very non trivial
verification that a Landauer Buttiker type approach
can safely  be applied to  integrable quasi particles.

To start, we recall the general expression for the current (\ref{curr})
\begin{equation}
I={1\over 2\pi}\int_{-\infty}^\infty d\theta {d \epsilon\over d\theta}
\left( {1\over 1+e^{(\epsilon-W/2)/T}}- {1\over
1+e^{(\epsilon+W/2)/T}}\right){1\over
1+e^{-2(t-1)(\theta-\theta_B)}}\label{wcur}
\end{equation}
where $\epsilon$ itself is a function of $T$. Recall also the value
\begin{equation}
e^{\epsilon(\theta=-\infty,T)/T}={\sinh (t-1)W/2tT \over \sinh W/2tT}
\end{equation}
We will only be interested in the terms of order $T$ and $T^2$ in the
current: we can therefore
drop exponentially small contributions, which makes matters considerably
simpler. For instance, only
the first term in (\ref{wcur}) contributes, and the value of
$\epsilon(-\infty,T)$ coincides at this order
with its value for $T=0$,
$\epsilon(-\infty,0)\equiv\epsilon_{min}={t-2\over 2t}W$.

To proceed, we consider the first term in (\ref{wcur}) and assume first
that $\epsilon(\theta)$ takes its $T=0$ value.
The finite $T$ corrections (we denote them by $\delta I^{(1)}$) then entirely
arise from a simple generalization of Sommerfeld's expansion in the
case of free electrons. We use here the same notations as in the appendix
of \cite{AM}. Introducing the function
\begin{equation}
H(\epsilon)={1\over 1+\left[{T_B\over e^{\theta(\epsilon)}}\right]^{2(t-1)}}
\end{equation}
we find
\begin{eqnarray}
I&=&{1\over 2\pi}{1\over
1+e^{(\epsilon_{min}-W/2)/T}}\int_{\epsilon_{min}}^{W\over 2} d\epsilon'
H(\epsilon')\nonumber\\
&+&{1\over 2\pi}\sum_{n=1}^\infty {1\over n!} {d^{n-1}\over
d\epsilon^{n-1}}\left. H\right|_{\epsilon=W/2}
\int_{\epsilon_{min}}^\infty \left(\epsilon-{W\over 2}\right)^n \times
-{d\over d\epsilon} {1\over 1+e^{(\epsilon-W/2)/T}}
\end{eqnarray}
Since we neglect exponentially small terms, we can neglect the filling
fraction in the first prefactor,
and replace the bound of integration in the integral by $-\infty$, using
the fact that $\epsilon_{min}<{W\over 2}$. It follows similarly
that only terms with $n$ {\sl even} contribute to the series, and therefore, to leading order,
\begin{equation}
\delta I^{(1)}={1\over 2\pi}\int_{\epsilon_{min}}^{W/2} d\epsilon'
H(\epsilon')+a_1{T^2\over 2\pi}{d\over d\epsilon}\left. H\right|_{\epsilon=W/2}
\end{equation}
The first term is nothing but $I(W,T=0)$. As for the second, $a_1$ is the
standard constant of the Sommerfeld expansion
\begin{equation}
a_1=\int_{-\infty}^\infty {\epsilon^2\over 2!}\times -{d\over d\epsilon}
{1\over 1+e^\epsilon} d\epsilon={\pi^2\over 6}
\end{equation}
At the order we are working, we finally obtain
\begin{equation}
\delta I^{(1)}(T)=T^2 {\pi\over 12} \left(\left.{d\epsilon\over
d\theta}\right|_{\theta=A}\right)^{-1} \left.{dH\over d\theta}\right|_{\theta=A}
\end{equation}
where $A$ is the Fermi momentum introduced in the text. One has on the other hand
$$
{dH\over d\theta}={t-1\over 2\cosh^2(t-1)(A-\theta_B)}
$$

To proceed, we must also take into account the changes of $\epsilon$ with
temperature  in the initial expression of the current.
The leading order correction turns out to be of order  $T^2$ then:
this gives a second contribution $\delta I^{(2)}$
to the change of the current,
and shows that   there are no crossed terms to this order.

Neglecting the exponentially small terms as before, the TBA equations for
$\epsilon$ do not need
the introduction of other pseudo energies and read
\begin{equation}
\epsilon(\theta)=e^\theta-T\int_{-\infty}^\infty \Phi(\theta-\theta')
\ln\left(1+e^{-(\epsilon(\theta')-W/2)/T}\right) d\theta'
\end{equation}
Integrations by part and Sommerfeld expansion give, as in the study of
$I^{(1)}$,
a leading correction going as $T^2$. We can thus write
$\epsilon(\theta,T)=\epsilon(\theta,T=0)+T^2\delta\epsilon$,
where we find
\begin{equation}
\delta\epsilon(\theta)-\int_{-\infty}^A
\phi(\theta-\theta')\delta\epsilon(\theta')d\theta'=- a_1T^2
\left(\left.{d\epsilon\over
d\theta}\right|_{\theta=A}\right)^{-1}
\phi(\theta-A)\label{inteq}
\end{equation}
This equation is solved by introducing the operator $L$ of \cite{LS}.
Calling the integral operator on the left of (\ref{inteq})
$\hat{I}-\hat{K}$ (where $\hat{I}$ is the identity),
one has $\hat{I}+\hat{L}={\hat{I}\over \hat{I}-\hat{K}}$. Using that
$\left(\hat{I}+\hat{K}\right)\bullet \phi=L$, it follows that
\begin{equation}
\delta\epsilon(\theta)=- a_1 T^2 \left(\left.{d\epsilon\over
d\theta}\right|_{\theta=A}\right)^{-1}L(\theta,A)
\end{equation}
Using the value
\begin{equation}
\left.{d\epsilon\over d\theta}\right|_A={W\over \sqrt{2t}}\label{resi}
\end{equation}
determined from \cite{LS}, we find therefore
\begin{equation}
\epsilon(\theta,T)=\epsilon(\theta,T=0)- T^2{\pi^2\over 3
W\sqrt{2/t}}L(\theta,A)
\end{equation}
Of course, the operator $L$ can be made explicit:
\begin{equation}
L(\theta,\theta')=L(\theta',\theta)=\phi(\theta-\theta')+\int_{-\infty}^A
\phi(\theta-\theta'')\phi(\theta''-\theta')d\theta''+\ldots
\end{equation}
The quantity $\epsilon$ we use here is related with another quantity
$\epsilon_+^h$   introduced in the main text (\ref{inteps}), and
studied in great details in \cite{LS}, by
$\epsilon={W\over 2}-\epsilon_+^h$. In the latter reference,
the following identity is established:
\begin{equation}
L(\theta,A)=-\sqrt{2t}W{d^2\epsilon\over dW^2}\label{resii}.
\end{equation}
Using this and integrating by parts, we find
\begin{equation}
\delta I^{(2)}(T)=T^2 {\pi\over 6 W}\sqrt{t\over 2}\int_{-\infty}^A d\theta
L(\theta,A) {dH\over d\theta}
\end{equation}
So collecting all terms,
\begin{equation}
I(T)=I(T=0)+ T^2 {\pi\over 6W}\sqrt{t\over 2} \left[\left. {dH\over
d\theta}\right|_{A} +\int_{-\infty}^A L(\theta,A)
 {dH\over d\theta}d\theta\right]\label{stepi}
\end{equation}
To conclude, we now turn to derivatives of the current with respect to $W$
at vanishing temperature. The current is usually written as
\begin{equation}
I(T=0)=\int_{-\infty}^A \rho(\theta) H(\theta)d\theta
\end{equation}
where the density $\rho$ is given by $\rho=-{1\over 2\pi}
{d\epsilon_+^h\over d\theta}$. Using integration by parts, one has
\begin{equation}
{dI\over dW}= {1\over 2\pi}\int_{-\infty}^A {d\epsilon_+^h\over dW}
{dH\over d\theta} d\theta
\end{equation}
Taking another derivative, using (\ref{resi}) and (\ref{resii}), one finds
\begin{equation}
{d^2 I\over dW^2}={1\over 2\pi W}\sqrt{1\over 2t}\left[\left. {dH\over
d\theta}\right|_A+\int_{-\infty}^A L(\theta,A) {dH\over d\theta}d\theta\right]
\end{equation}
and thus, comparing with (\ref{stepi})
\begin{equation}
I(W,T)=I(W,T=0)+t{\pi^2 T^2 \over 3} {d^2 I\over dW^2}(W,T=0)
\end{equation}
(this, up to exponentially small terms and higher order analytical terms),
thus proving the identity.

As commented in the main text and in \cite{FLS}, the differential conductance
for $g<{1\over 2}$ is {\sl negative} for large enough $W/T_B$ (this result
does not rely on the Bethe ansatz, and is a simple consequence of the
non linear $I-W$ curve present in the Luttinger liquid). It follows
from (\ref{weissrel}) that for such values of $g$, the current in the fractional quantum Hall problem {\sl diminishes}
 when $T$ is increased from $T=0$, provided $W/T_B$ is large enough. This is
a rather counterintuitive phenomenon: a priori, one expects that, the larger $T$, the more energy there is, and therefore the less important the backscattering should be. Of course, the current depends on more complex details than the
overall energy, and it is well possible that $W,T$, and the non trivial interactions produce an overall less efficient population of quasi-particles, even though $T$ is increased. Notice that the current can also decrease when $T$ is turned on now at fixed $U$, as is clear on figure 5.

To conclude, observe that, using (\ref{weissrel}) together with the
duality relation at $T=0$,
the same relation is found to hold to order $T^2$,  in agreement
with the fact that the duality relation should actually hold
at any temperature.

\end{document}